\documentclass{PoS}

\usepackage{amsmath,amssymb}
\usepackage{graphicx}
\usepackage{subfigure}

\title{
\vspace{-1cm}
{\small \normalfont 
  \hfill DESY 12-214 \\
  \hfill SFB/CPP-12-90\\
  \hfill HU-EP-12/48 \\
}
\vspace{0.2cm}
{Properties of Pseudoscalar Flavour-Singlet Mesons from $N_f=2+1+1$ Twisted Mass Lattice QCD}}

\ShortTitle{$\eta$, $\eta'$ meson masses}

\author{Krzysztof Cichy\\
  NIC, DESY, Platanenallee 6, D-15738 Zeuthen, Germany\\
  \& Adam Mickiewicz University, Faculty of Physics,
  Umultowska 85, 61-614 Poznan, Poland\\
  \email{krzysztof.cichy@desy.de}
}

\author{Vincent Drach, Karl Jansen\\
  NIC, DESY, Platanenallee 6, D-15738 Zeuthen, Germany\\
  \email{vincent.drach,karl.jansen@desy.de}
}

\author{Elena Garcia-Ramos\\
  NIC, DESY, Platanenallee 6, D-15738 Zeuthen, Germany\\
  \& Humboldt Universit\"at zu Berlin, Newtonstrasse 15, D-12489 Berlin, Germany\\
  \email{elena.garcia.ramos@desy.de}
}

\author{Chris Michael\\
  Theoretical Physics Division, Department of Mathematical Sciences,\\
  The University of Liverpool, Liverpool L69 3BX, UK\\
  \email{C.Michael@liverpool.ac.uk}
}

\author{\speaker{Konstantin Ottnad}, Carsten Urbach, \speaker{Falk Zimmermann}\\
  Institut f{\"u}r Strahlen- und Kernphysik, Rheinische Friedrich-Wilhelms-Universit{\"a}t Bonn\\
  Nussallee 14-16, 53115 Bonn \\ 
  \email{ottnad,urbach,fzimmermann@hiskp.uni-bonn.de}
}

\author{For the ETM Collaboration}

\abstract{We study properties of pseudoscalar flavour-singlet mesons
  from Wilson twisted mass lattice QCD with $N_f=2+1+1$ dynamical
  quark flavors. Results for masses are presented at three values of
  the lattice spacing and 
  light quark masses corresponding to values of the pion mass from
  $230\, \mbox{MeV}$ to $500\,\mbox{MeV}$. We briefly discuss
  scaling effects and the light and strange quark mass dependence of
  $M_\eta$. In addition we present an exploratory study using
  Osterwalder-Seiler type strange and charm valence quarks. This
  approach avoids some of the 
  complications of the twisted mass heavy doublet. We present first
  results for matching valence and unitary actions and a comparison of
  statistical uncertainties.}

\FullConference{The 30 International Symposium on Lattice Field Theory - Lattice 2012,\\
		June 24-29, 2012\\
		Cairns, Australia}

\begin{document}

\maketitle






\section{Introduction}

A remarkable feature of the $\eta'$ meson in comparison to the pions,
the kaons and the $\eta$ meson is its extraordinary high mass of about
$1$ GeV: while pions consist only of light quarks and - therefore -
exhibit a rather small mass around $140$ MeV, strange quark
contributions rise the mass of the four kaons and the $\eta$ meson to
$500$ - $600$ MeV. However, the larger strange quark mass cannot
explain the mass value of the $\eta'$ meson. On the QCD level, the
reason for the mass gap is thought to be the breaking of the $U_A(1)$
symmetry by quantum effects. 

Although lattice QCD provides access to flavour singlet pseudo-scalar
states, significant contributions from quark disconnected diagrams
complicate the determination of their properties. This might be the
reason for the rather short list of publications covering these
mesons. Recent studies for $2+1$ dynamical quark flavours can be 
found in \cite{Christ:2010dd,Kaneko:2009za,Dudek:2011tt} and
\cite{Gregory:2011sg}. For a recent publication with $N_f=2$ Wilson
twisted mass fermions see Ref.~\cite{Jansen:2008wv}.

In this proceeding contribution, we build on a recent paper of some of
the authors~\cite{Ottnad:2012fv} and discuss the determination of 
$\eta$ and $\eta'$ meson masses and the mixing angle using the Wilson
twisted mass lattice QCD formulation \cite{Frezzotti:2000nk} with
$N_f=2+1+1$ dynamical quark flavours. Compared to
Ref.~\cite{Ottnad:2012fv} we present results for an additional
ensemble, which allows us to better control
systematic uncertainties. The final results stay virtually
unaffected compared to Ref.~\cite{Ottnad:2012fv}. 

The results presented in Ref.~\cite{Ottnad:2012fv} were extracted for
the unitary case, with identical sea and valence quark
regularisations. In addition we discuss a mixed action 
approach with so called Osterwalder-Seiler strange and
charm valence quarks~\cite{Frezzotti:2004wz}. Possible advantages of
such an approach are a powerful variance reduction technique for
strange and charm quarks~\cite{Jansen:2008wv,Dinter:2012tt} and
reduced twisted mass induced isospin splitting in the strange/charm
doublet. We discuss how to perform the matching of valence and sea
formulation and present first results for the variance reduction.


\section{Lattice actions}
\label{sec:actions}

The results we present are based on gauge configurations provided by
the European Twisted Mass Collaboration (ETMC) and correspond to three
values of the lattice spacing, $a=0.061$ fm, $a=0.078$ fm and
$a=0.086$ fm. The pion masses range from $230$ to $500$
MeV~\cite{Baron:2010bv,Baron:2011sf}. A list of the
investigated ensembles is given in Table~\ref{tab:setup}. For setting
the scale we use throughout this proceeding contribution the Sommer
parameter $r_0 = 0.45(2)\ \mathrm{fm}$~\cite{Baron:2011sf}.

The Dirac operators for $u$ and $d$ quarks~\cite{Frezzotti:2000nk} reads
\begin{equation}
 D_\ell = D_W + m_0 + i \mu_\ell \gamma_5\tau^3
 \label{eq:Dlight}
\end{equation}
with $\mu_\ell$ the bare light twisted mass parameter.
For the heavy doublet of $c$ and $s$ quarks~\cite{Frezzotti:2003xj}
the Dirac operator reads
\begin{equation}
 D_h = D_W + m_0 + i \mu_\sigma \gamma_5\tau^1 + \mu_\delta \tau^3\,.
 \label{eq:Dsc}
\end{equation}
$D_W$ denotes the standard Wilson operator and the $\tau^i$ are Pauli
matrices acting in flavour space. The value of
the bare quark mass $m_0$ was tuned to its critical
value~\cite{Chiarappa:2006ae,Baron:2010bv}, leading to automatic order
$\mathcal{O}\left(a\right)$ improvement at maximal twist
\cite{Frezzotti:2003ni}, which represents the most notable advantage
of tmLQCD. The two bare quark masses $\mu_\sigma$, $\mu_\delta$
are related to the physical charm and strange quark mass via 
\begin{equation}
 m_{c,s} = \mu_\sigma\ \pm\ Z\ \mu_\delta \, ,
 \label{eq:msc}
\end{equation}
where $Z=Z_\mathrm{P}/Z_\mathrm{S}$ defines the ratio of pseudo-scalar
and scalar renormalisation constants $Z_P$ and $Z_S$. We denote quark
doublets in the physical basis by $\psi_{\ell,h}$ and in the twisted
basis by $\chi_{\ell,h}$. In the continuum they are related by exact
axial rotations 
\begin{equation}
 \psi_{\ell,h} = e^{i \pi \gamma_5 \tau^{3,1} / 4} \, \chi_{\ell,h} \,
 , \ \ \ \bar{\psi}_{\ell,h} = \bar{\chi}_{\ell,h} \, e^{i \pi
   \gamma_5 \tau^{3,1} / 4} \, . 
 \label{eq:unitaryfields}
\end{equation}
The main drawback of this formulation is the breaking of flavour
symmetry at finite values of the lattice spacing, which was shown to
affect mainly the value of the neutral pion
mass~\cite{Urbach:2007rt,Dimopoulos:2009qv,Baron:2009wt}. Furthermore,
for the non-degenerate quark doublet this introduces mixing between
charm and strange quarks.

This complication can be circumvented in a mixed action approach using
so-called Osterwalder-Seiler (OS) type quarks in the valence sector for
strange and charm quarks~\cite{Frezzotti:2004wz}. Formally, we
introduce one twisted doublet each for valence strange and valence
charm quarks~\cite{Frezzotti:2004wz,Blossier:2007vv}. Therefore, in
the valence strange and charm sector flavour mixing is avoided. It was
shown in Ref.~\cite{Frezzotti:2004wz} that automatic $\mathcal{O}(a)$
improvement is not spoiled by this approach and that unitarity is
restored in the continuum limit.

Formally, we have introduced two strange (charm) quarks differing in
the sign of the twisted mass. We will denote the one with positive sign
with $s$ ($c$) and the other one with $s'$ ($c'$). With $\mu_s$ and
$\mu_c$ we denote  the bare OS strange and charm twisted masses,
respectively.

The two actions in the sea and the valence sector must be matched
appropriately. This matching can be performed using
different observables: firstly, one can match using the quark mass
values
\[
\mu_{s,c} = \mu_\sigma\ \pm\ Z\ \mu_\delta\,,
\]
requiring the knowledge of the ratio of renormalisation constants
$Z$; secondly, unitary and OS kaon mass values can be matched. For the
OS kaons we have the possibility to define the following OS
interpolating fields in the twisted basis
\[
\mathcal{O}_{K^{\mathrm{OS}}}\equiv \bar\chi_{s'}\, \chi_d(x)\,,
\qquad \mathcal{O}_{K^{+}}\equiv \bar\chi_s\, i\gamma_5 \chi_d\,,
\]
which will lead to different kaon mass values at finite values of the
lattice spacing. The corresponding kaon masses will be denoted by
$M_{\mathrm{K}^\mathrm{OS}}$ and $M_{\mathrm{K}^+}$, respectively. The
unitary kaon mass value is denoted by $M_\mathrm{K}$. Note that there
is no isospin splitting for the kaons in $N_f=2+1+1$ Wilson twisted
mass lattice QCD \cite{Baron:2010th}.

Thirdly, one can use the mass of the artificial $\eta_s$ meson -- a
pion made out of strange quarks -- to  
match sea and valence actions. $M_{\eta_s}$ can be determined from the
connected only correlation function of the following operator
\[
\mathcal{O}_{\eta_s}\ \equiv\ \frac{1}{\sqrt{2}}(\bar\chi_s\, \chi_s - \bar
\chi_{s'}\,\chi_{s'})\,.
\]

Similarly, one can define matching observables for the charm
quark. However, as we are mainly interested in $\eta$ and $\eta'$
mesons, we expect little impact from the charm quark. Therefore, we
will not investigate different matching observables for the charm quark and
use only one $\mu_c$-value. It is obtained by matching $aM_{\mathrm{K}^+}
=aM_\mathrm{K}$ to determine $a\mu_s$. The latter is
then used to obtain $a\mu_c$ via the relation Eq.~(\ref{eq:msc}).

For the extraction of quark masses and pseudo-scalar decay constants
it turned out that using $M_{\mathrm{K}^+}$ for the matching lead to the
smallest lattice artifacts~\cite{Sharpe:2004ny}. However, this may depend on the quantities
under consideration.

\begin{table}[t!]
 \centering
 \begin{tabular*}{1.\textwidth}{@{\extracolsep{\fill}}lcccccccc}
  \hline\hline
  ensemble & $\beta$ & $a\mu_\ell$ & $a\mu_\sigma$ & $a\mu_\delta$ & $L/a$ & $N_\mathrm{conf}$ & $N_s$ & $N_b$ \\ 
  \hline\hline
  $A30.32$   & $1.90$ & $0.0030$ & $0.150$  & $0.190$  & $32$ & $1367$ & $24$ & $5$  \\
  $A40.24$   & $1.90$ & $0.0040$ & $0.150$  & $0.190$  & $24$ & $2630$ & $32$ & $10$ \\
  $A40.32$   & $1.90$ & $0.0040$ & $0.150$  & $0.190$  & $32$ & $863$  & $24$ & $4$  \\
  $A60.24$   & $1.90$ & $0.0060$ & $0.150$  & $0.190$  & $24$ & $1251$ & $32$ & $5$  \\
  $A80.24$   & $1.90$ & $0.0080$ & $0.150$  & $0.190$  & $24$ & $2449$ & $32$ & $10$ \\
  $A100.24$  & $1.90$ & $0.0100$ & $0.150$  & $0.190$  & $24$ & $2493$ & $32$ & $10$ \\
  \hline
  $A80.24s$  & $1.90$ & $0.0080$ & $0.150$  & $0.197$  & $24$ & $2517$ & $32$ & $10$ \\
  $A100.24s$ & $1.90$ & $0.0100$ & $0.150$  & $0.197$  & $24$ & $2312$ & $32$ & $10$ \\
  \hline
  $B25.32$   & $1.95$ & $0.0025$ & $0.135$  & $0.170$  & $32$ & $1484$ & $24$ & $5$  \\
  $B35.32$   & $1.95$ & $0.0035$ & $0.135$  & $0.170$  & $32$ & $1251$ & $24$ & $5$  \\
  $B55.32$   & $1.95$ & $0.0055$ & $0.135$  & $0.170$  & $32$ & $1545$ & $24$ & $5$  \\ 
  $B75.32$   & $1.95$ & $0.0075$ & $0.135$  & $0.170$  & $32$ & $922$  & $24$ & $4$  \\ 
  $B85.24$   & $1.95$ & $0.0085$ & $0.135$  & $0.170$  & $24$ & $573$  & $32$ & $2$  \\
  \hline
  $D15.48$   & $2.10$ & $0.0015$ & $0.120$  & $0.1385$ & $48$ & $1045$ & $24$ & $10$ \\
  $D30.48$   & $2.10$ & $0.0030$ & $0.120$  & $0.1385$ & $48$ & $469$  & $24$ & $3$  \\
  $D45.32sc$ & $2.10$ & $0.0045$ & $0.0937$ & $0.1077$ & $32$ & $1887$ & $24$ & $10$ \\
  \hline\hline
  \vspace*{0.1cm}
 \end{tabular*}
 \caption{The ensembles used in this investigation. For the labeling
   we employ the notation of ref.~\cite{Baron:2010bv}. Additionally,
   we give the number of configurations $N_\mathrm{conf}$, the number
   of stochastic samples $N_s$ for all ensembles and the bootstrap
   block length $N_b$. The D30.48 ensemble was not yet included in
   Ref.~\cite{Ottnad:2012fv}.}
 \label{tab:setup}
\end{table}

\section{Pseudoscalar flavour-singlet mesons}
In order to study properties of pseudoscalar flavour-singlet mesons
we have to consider light strange and charm contributions to build a
suitable correlation function matrix. Hence, in the physical basis we
are after computing 
\begin{equation}
  \label{eq:corrmatrix}
  \mathcal{C}(t) = 
  \begin{pmatrix}
    \eta_\ell(t)\eta_\ell(0) & \eta_\ell(t)\eta_s(0) & \eta_\ell(t)\eta_c(0) \\
    \eta_s(t)\eta_\ell(0) & \eta_s(t)\eta_s(0) & \eta_s(t)\eta_c(0) \\
    \eta_c(t)\eta_\ell(0) & \eta_c(t)\eta_s(0) & \eta_c(t)\eta_c(0) \\
  \end{pmatrix}\,,
\end{equation}
involving the following interpolating operators
\begin{equation}
  \label{eq:operators}
  \eta_\ell\equiv(\bar\psi_u i\gamma_5 \psi_u+\bar\psi_d i\gamma_5 \psi_d)/\sqrt{2},\quad
  \eta_s\equiv(\bar\psi_s i\gamma_5 \psi_s),\quad \eta_c\equiv(\bar
  \psi_ci\gamma_5 \psi_c)\, . 
\end{equation}
Since we are working in the twisted basis we have to rotate these
operators. Let us start with the unitary action, where the rotation in
the light sector is given by~\cite{Jansen:2008wv} 
\begin{equation}
 \frac{1}{\sqrt{2}}(\bar{\psi}_u i \gamma_5\psi_u + \bar{\psi}_d i
 \gamma_5\psi_d) \quad \to \quad \frac{1}{\sqrt{2}}(- \bar\chi_u
 \chi_u + \bar\chi_d \chi_d)\ \equiv\ \mathcal{O}_{\ell}\, , 
 \label{eq:Oplight}
\end{equation}
where the left- and right-hand side correspond to physical and twisted
basis, respectively. Also for strange and charm we have to consider a
doublet as follows
\begin{equation}
 \begin{pmatrix}
  \bar \psi_c \\
  \bar \psi_s \\
 \end{pmatrix}^T
 i \gamma_5 \frac{1 \pm \tau^3}{2}
 \begin{pmatrix}
  \psi_c \\
  \psi_s \\
 \end{pmatrix}
 \quad\to\quad
 \begin{pmatrix}
  \bar \chi_c \\
  \bar \chi_s \\
 \end{pmatrix}^T
 \frac{-\tau^1  \pm  i \gamma_5\tau^3}{2}
 \begin{pmatrix}
  \chi_c \\
  \chi_s \\
 \end{pmatrix}\ \equiv\ \mathcal{O}_{c,s}\, .
 \label{eq:Opheavy}
\end{equation}
The flavour space projector $(1\pm\tau^3)/2$ distinguishes between charm
and strange contributions in the physical basis. In the twisted basis
we need to consider the following operators when calculating
correlation functions 
\begin{equation}
 \begin{split}
  \mathcal{O}_{c}\ &\equiv\ Z(\bar\chi_c i \gamma_5\chi_c - \bar\chi_s
  i \gamma_5\chi_s)/2 - (\bar\chi_s\chi_c + \bar\chi_c\chi_s)/2\, ,\\   
  \mathcal{O}_{s}\ &\equiv\ Z(\bar\chi_s i \gamma_5\chi_s - \bar\chi_c
  i \gamma_5\chi_c)/2 - (\bar\chi_s\chi_c + \bar\chi_c\chi_s)/2\, .\\ 
 \end{split}
 \label{eq:OpheavyTm}
\end{equation}
Note that the ratio of renormalisation constants $Z$ appears in the
sum of pseudoscalar and scalar currents. However, $Z$ is not needed
for extracting the masses of $\eta$ and $\eta'$, as we explain in
Ref.~\cite{Ottnad:2012fv}. For further details on how to construct the
correlation matrix we also refer to Ref.~\cite{Ottnad:2012fv}. 

Let us now discuss the rotation for the mixed action approach with OS
valence strange and charm quarks. For the light quarks the operator is
identical to the unitary one in Eq.~(\ref{eq:Oplight}),
i.e. $\mathcal{O}^\mathrm{OS}_\ell \equiv \mathcal{O}_\ell$. But for
strange and charm quarks the operators are significantly simpler, due
to no flavour mixing in between strange and charm:
\begin{equation}
  \label{eq:OSop}
  \begin{split}
    &\frac{1}{\sqrt{2}}(\bar{\psi}_c i
    \gamma_5 \psi_c \ + \ \bar{\psi}_{c'} i \gamma_5 \psi_{c'}) ~ \to ~
    \frac{1}{\sqrt{2}}(\bar{\chi}_c \chi_c \ - \ \bar{\chi}_{c'}
    \chi_{c'})\ \equiv\mathcal{O}_{c}^\mathrm{OS}\,, \\ 
    &\frac{1}{\sqrt{2}}(\bar{\psi}_s i
    \gamma_5 \psi_s \ + \ \bar{\psi}_{s'} i \gamma_5 \psi_{s'}) ~ \to ~
    \frac{1}{\sqrt{2}}(\bar{\chi}_s \chi_s \ - \ \bar{\chi}_{s'}
    \chi_{s'})\ \equiv\ \mathcal{O}_{s}^\mathrm{OS}\,.
  \end{split}
\end{equation}
Again, we build a correlation function matrix of the form given in
Eq.~(\ref{eq:corrmatrix}). Obviously, there is no mixing of pseudoscalar
and scalar currents like in Eq.~(\ref{eq:OpheavyTm}) and,
therefore, no ratio of renormalisation constants appears.

For both, the unitary and the mixed action approach we solve the
generalised eigenvalue
problem~\cite{Michael:1982gb,Luscher:1990ck,Blossier:2009kd}  
\begin{equation}
 \mathcal{C}(t)\ \eta^{(n)}(t,t_0) = \lambda^{(n)}(t, t_0)\
 \mathcal{C}(t_0)\ \eta^{(n)}(t,t_0) \, .
 \label{eq:gevp}
\end{equation}
Taking into account the periodic boundary conditions for a meson and
solving 
\begin{equation}
 \frac{\lambda^{(n)}(t,t_0)}{\lambda^{(n)}(t+1,t_0)} =
 \frac{e^{-m^{(n)} t}+ e^{-m^{(n)}(T-t)}} {e^{-m^{(n)}(t+1)}+
   e^{-m^{(n)}(T-(t+1))}} 
\end{equation}
we determine the effective masses $m^{(n)}$, where $n$ counts the
eigenvalues. The state with the lowest mass should correspond to the
$\eta$ and the second state to the $\eta'$ meson. Alternatively, we
use a factorising fit of the form  
\begin{equation}
 \mathcal{C}_{qq'}(t) = \sum_n \frac{A_{q,n}A_{q',n}}{2 m^{(n)}}\
 \left[\exp(-m^{(n)} t) + \exp(-m^{(n)}(T-t))\right] 
 \label{eq:fit}
\end{equation}
to the correlation matrix matrix $\mathcal{C}$. The amplitudes
$A_{q,n}$ correspond to $\langle0|\bar q q|n \rangle$ with $n\equiv
\eta,\eta',...$ and $q = \ell, s, c$. Note that for this physical
interpretation of the amplitudes the ratio of renormalisation
constants $Z$ is unavoidably required as input for the unitary
approach~\cite{Ottnad:2012fv}.

\subsection{Variance reduction}
\label{sec:var}
In general the correlation functions consist of quark connected and
disconnected diagrams. The connected pieces have been calculated via
the so called  ``one-end-trick'' \cite{Boucaud:2008xu} using
stochastic timeslice sources. For the disconnected diagrams we resort
to stochastic volume sources with complex Gaussian noise. The light
disconnected contributions can be estimated very efficiently using the
identity \cite{Jansen:2008wv} 
\begin{equation}
  \label{eq:vv}
  D_u^{-1} - D_d^{-1} = -2i\mu_\ell D_d^{-1}\ \gamma_5\ D_u^{-1}\ .
\end{equation}
In the heavy sector of the unitary setup such a simple identity does
not exist. Instead we use the (less efficient) so called hopping
parameter variance reduction, which relies on the same identity as in
the mass degenerate two flavour case (see ref.~\cite{Boucaud:2008xu}
and references therein).
The number of stochastic volume sources $N_s$ per gauge configuration
we used for both the heavy and the light sector is given for each
ensemble in table~\ref{tab:setup}. In order to check that the
stochastic noise introduced by our method is smaller than the gauge
noise we have increased $N_s$ from $24$ to $64$ for ensemble $B25.32$,
which did not reduce the error on the extracted masses. 

For OS strange and charm quarks the variance reduction trick
Eq.~(\ref{eq:vv}) is also applicable, as noted in
Ref.~\cite{Dinter:2012tt}. Consider the disconnected contributions for
the operator in Eq.~(\ref{eq:OSop}). Again, we may write for the
strange quark
\begin{equation}
  \label{eq:OSvv}
  D_s^{-1} - D_{s'}^{-1} = -2i\mu_s D_{s'}^{-1}\ \gamma_5\ D_s^{-1}\,,
\end{equation}
and similarly for the charm quark. This identity can be used like in the
light sector to compute the disconnected contributions of strange and
charm quarks to the OS correlator matrix with greatly reduced
noise~\cite{Dinter:2012tt}.

\section{Results from Unitary Strange and Charm Quarks}

Most of the unitary results have already been published
in Ref.~\cite{Ottnad:2012fv}, the only exception is ensemble $D30.48$.
However this additional point does only very mildly affect
the final results.

We have calculated all required contractions for the correlator
matrix using local and fuzzed operators, yielding
a $6 \times 6$-matrix. The number of gauge configurations per ensemble
is given in Table~\ref{tab:setup}. All errors have been calculated
from bootstrapping with $1000$ samples. To compensate for
autcorrelation we have used blocking, the number of configuration per
block $N_b$ is also given in Table~\ref{tab:setup} and it was chosen 
such that the resulting blocklength in HMC trajectories
$N_{\text{HMC}}$ fulfils $N_{\text{HMC}}\geq20$. For a more detailed
discussion on autocorrelation, which mainly affects the $\eta'$-state
we refer to \cite{Ottnad:2012fv}.

\subsection{Extraction of Masses}

\begin{figure}[t]
 \centering
 \subfigure[]{\includegraphics[width=.47\linewidth]{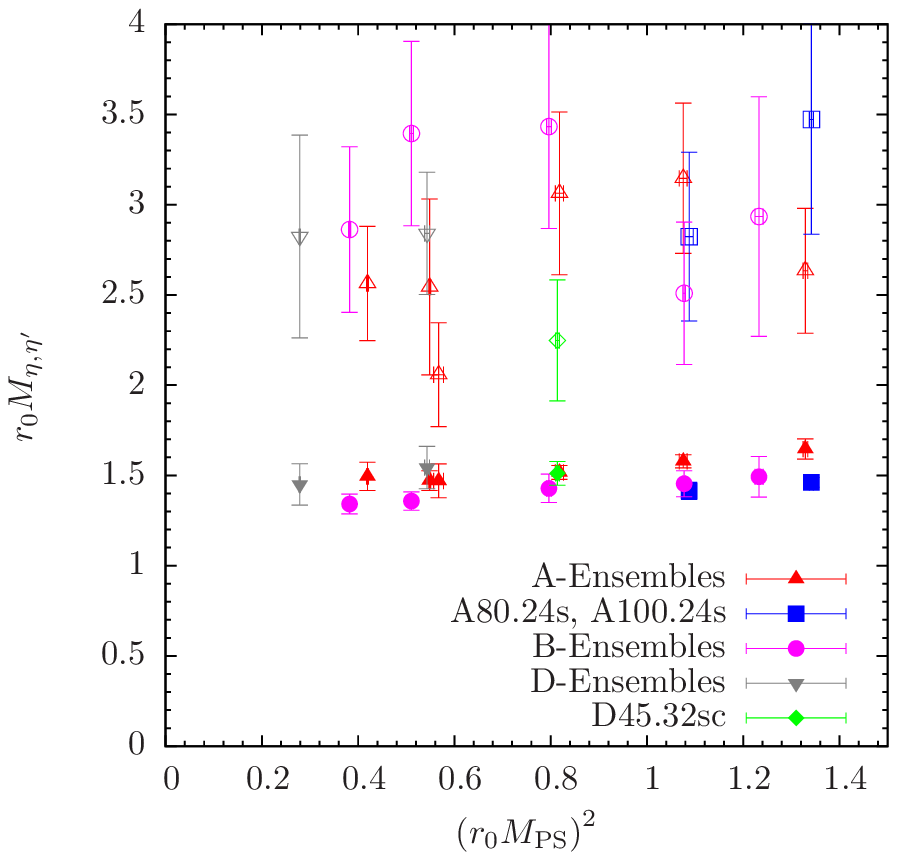}}\quad
 \subfigure[]{\includegraphics[width=.49\linewidth]{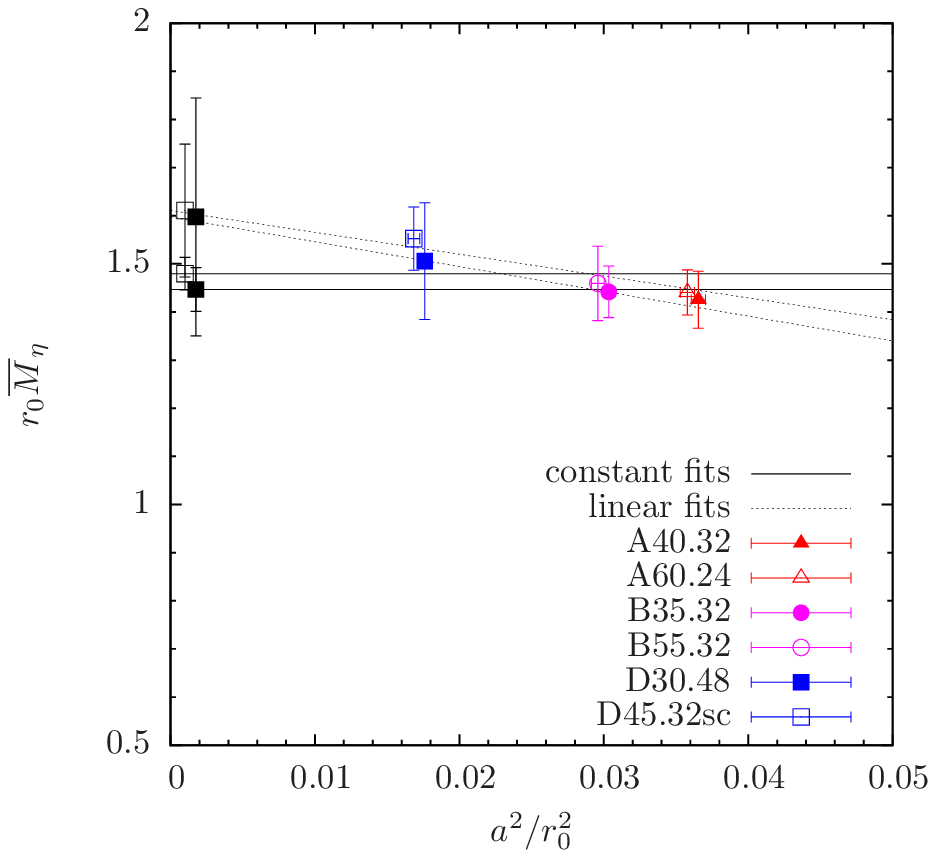}}\quad
 \caption{(a) $\eta$ (filled symbols) and $\eta'$ (open symbols)
   masses in units of $r_0$. (b) $r_0\overline{M}_\eta$ as a function
   of $(a/r_0)^2$ for the ensemble sets $S_1$ and $S_2$.} 
 \label{fig:masses}
\end{figure}

The details of our GEVP and fitting procedures to extract $\eta$ and
$\eta'$ masses are explained in Ref.~\cite{Ottnad:2012fv}. In
Figure~\ref{fig:masses} we show the masses of the $\eta$ (filled
symbols) and $\eta'$ (open symbols) 
mesons for the various ensembles we used as a function of the squared
pion mass, everything in units of $r_0$. The values of the chirally
extrapolated $r_0^\chi$ for each value of $\beta$ can also be found in
Ref.~\cite{Ottnad:2012fv}. We present the values for
$aM_\eta$ and $aM_{\eta'}$ together with kaon and pion mass values in
Table~\ref{tab:masses} with statistical errors only for the two
ensembles $D30.48$ and $D45.32sc$. The results for $D30.48$ are new
compared to what was shown in Ref.~\cite{Ottnad:2012fv}. $D45.32sc$
will be used for the mixed action analysis. All other results can be
found in Ref.~\cite{Ottnad:2012fv}.

It is clear from the figure that the $\eta$ meson mass can be
extracted with high precision, while the $\eta'$ meson mass is more
noisy.

\subsection[Scaling Artifacts and Strange Quark Mass Dependence of the
eta mass]{Scaling Artifacts and Strange Quark Mass Dependence of
  $M_\eta$} 

The results displayed in the left panel of Figure~\ref{fig:masses}
have been obtained using the bare values of $a\mu_\sigma$ and
$a\mu_\delta$ as used for the production of the ensembles. Those
values, however, did not lead to the physical values of, e.g., the
kaon and D-meson masses~\cite{Baron:2010bv,Baron:2010th}. 

For $M_\eta$ the statistical uncertainty is sufficiently small to
attempt to correct for the mismatch in the strange quark mass value
and to try a scaling test. For this we need to compare $M_\eta$ at the
three different values of the lattice spacing for fixed values of for
instance $r_0 M_\mathrm{K}$, $r_0M_\mathrm{D}$, $r_0M_\mathrm{PS}$ and
the physical volume. From volume and the charm quark mass value we
expect only little influence given our uncertainties and hence, we are
going to disregard these minor effects in the following.  

As discussed in Ref.~\cite{Ottnad:2012fv}, we have to perform
an interpolation of $M_\eta$ in $M_\mathrm{K}$. For this purpose, we
treat the masses of the $\eta$-meson and the kaon like in chiral
perturbation theory as functions $M^2 = M^2[M_\mathrm{PS}^2,
M_\mathrm{K}^2]$ and define the dimensionless derivative
\begin{equation}
  \label{eq:Deta}
  D_\eta(\mu_\ell, \mu_\sigma, \mu_\delta, \beta)\ \equiv\
  \left[\frac{d (a M_\eta)^2}{d (a M_\mathrm{K})^2}\right]\ .
\end{equation}
Next we make the approximation that $D_\eta$ is independent of the
quark mass values $\mu_\ell, \mu_\sigma, \mu_\delta$ and $\beta$. Its
value we can estimate from $A80.24$ and
$A80.24s$ as well as from $A100.24$ and $A100.24s$. On average we obtain
$D_\eta =1.60(18)$.  

Now we use this value of $D_\eta$ to correct two sets $S_1, S_2$ of
three ensembles, namely $S_1 = \left\{A40.32, B35.32, D30.48\right\}$
and $S_2=\left\{A60.24, B55.24, D45.24\right\}$  to a common value of
$r_0 M_\mathrm{K}\approx1.34$ using 
\[
(r_0 \overline{M}_\eta)^2 = (r_0 M_\eta)^2 +
  D_\eta\cdot\Delta_\mathrm{K}\, ,
\]
where $\Delta_\mathrm{K}$ is the difference in the squared kaon mass
values to the squared reference values (in units of $r_0$). For each
set the three points have approximately fixed values of $r_0
M_\mathrm{PS}$.

\begin{figure}[t]
 \centering
 \subfigure[]{\includegraphics[width=.48\linewidth]{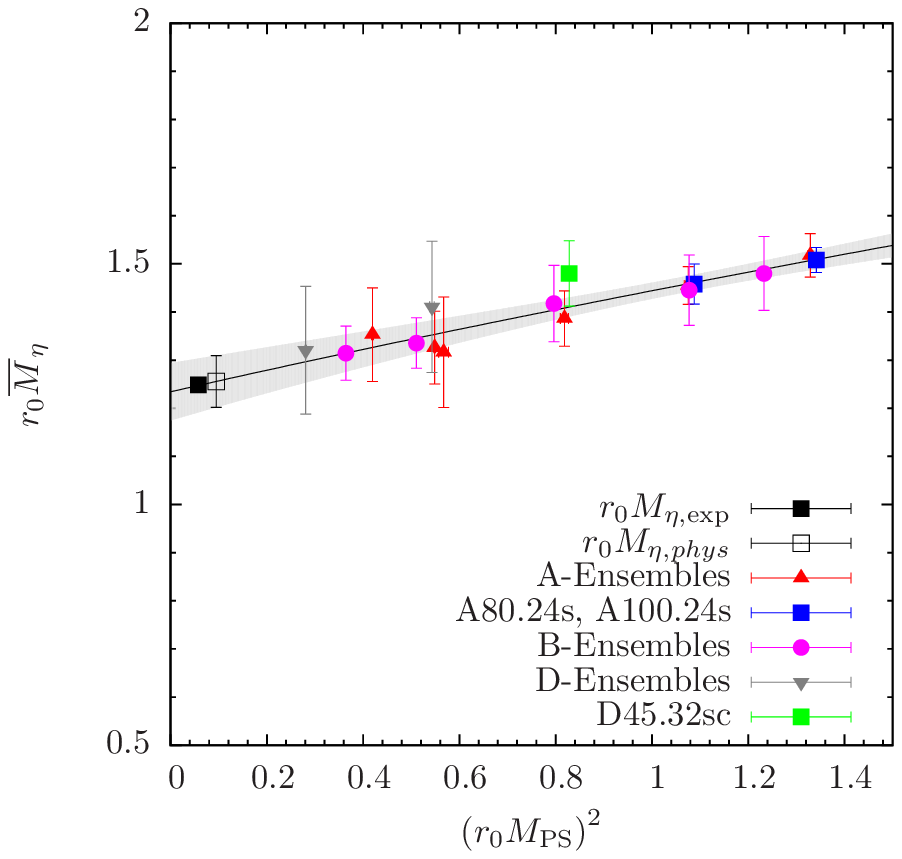}}
 \subfigure[]{\includegraphics[width=.48\linewidth]{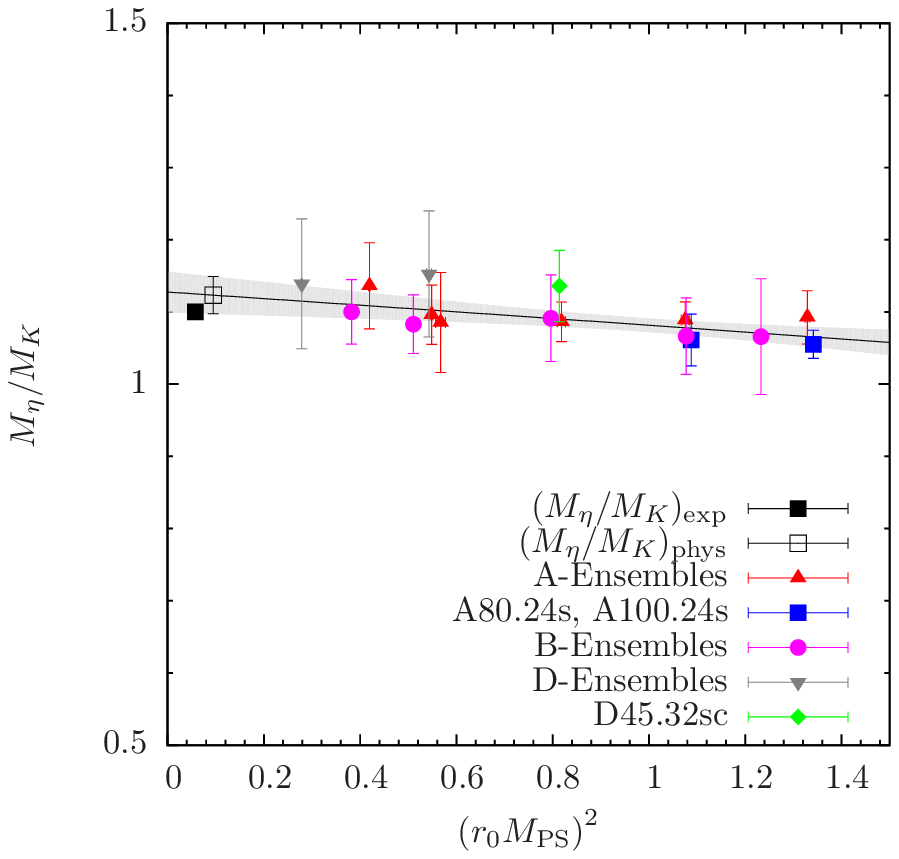}}
 \caption{(a) $r_0 \overline{M}_\eta$ as a function of
   $(r_0M_\mathrm{PS})^2$. (b) $M_\eta/M_\mathrm{K}$ as a function of $(r_0M_\mathrm{PS})^2$.} 
 \label{fig:r0meta}
\end{figure}

We plot the resulting $r_0 \overline{M}_\eta$ values for sets
$S_{1,2}$ as a function of $(a/r_0)^2$ in the right panel of
Figure~\ref{fig:masses}. Both data sets are still
compatible with a constant continuum extrapolation giving $r_0
M_{\eta,S_1,\mathrm{const}}^{ \rightarrow0} = 1.447(45)$ and $r_0
M_{\eta,S_2,\mathrm{const}}^{ \rightarrow0} = 1.480(34)$,
respectively, which we indicate by the horizontal lines. We can also
perform a linear extrapolation, leading to $r_0
M_{\eta,S_1,\mathrm{lin}}^{a\rightarrow0} = 1.60(25)$ and $r_0
M_{\eta,S_2,\mathrm{lin}}^{a\rightarrow0} = 1.61(14)$, which is also
shown in the figure. The difference in between the two extrapolated
values for each set are 
\begin{equation}
r_0 \Delta M_{\eta,S_1}^{a\rightarrow0} = 0.15(25) \ , \quad r_0
\Delta M_{\eta,S_2}^{a\rightarrow0} = 0.13(13)  
\label{eq:deltascaling}
\end{equation}
and they give us an estimate on the systematic uncertainty to be
expected from the continuum extrapolation. Both results agree well, although 
the one for $S_1$ exhibits twice the error.
We will therefore quote an $8\%$ relative error from $\Delta
M_{\eta,S_2}^{a\rightarrow0}/M_{\eta,S_2,
  \mathrm{const}}^{a\rightarrow0}$ for our mass estimates, which was
already used in Ref.~\cite{Ottnad:2012fv} where $S_1$ was not yet
available.

In order to obtain a more complete picture, we now correct all our
ensembles for the slightly mistuned value of 
$M_\mathrm{K}$. For this we follow the procedure which was discussed
in detail in \cite{Ottnad:2012fv}, i.e. we shift the kaon mass values
for all ensembles to a common line $(r_0 M_K)^2[(r_0 M_\mathrm{PS})^2]$
determined such that it reproduces the physical kaon mass value at the
physical point. Next we correct the $\eta$ masses appropriately. The
result of this procedure is 
shown in the left panel of Figure~\ref{fig:r0meta}: we show values of
the corrected $\eta$ masses $r_0 \overline{M}_\eta$ for all our ensembles
as a function of $(r_0 M_\mathrm{PS})^2$. It is evident that all the
data fall on a single curve within statistical uncertainties, which
confirms that $M_\eta$ is not affected by large cut-off effects. Note,
however, that we again ignored possible $\mu_\ell$, $\mu_\sigma$,
$\mu_\delta$ and $\beta$ dependence with this procedure. 

\subsection{Extrapolation to the Physical Point}

Since we have now fixed the strange quark mass to its physical value
using $M_\mathrm{K^0}^\mathrm{exp} = 498\ \mathrm{MeV}$, we can attempt
a linear fit to all corrected data points for
$(r_0\overline{M}_\eta)^2[(r_0 M_\mathrm{PS})^2]$. Using $r_0=0.45(2)\,\mathrm{fm}$
as in Ref.~\cite{Baron:2011sf}, the fit yields $r_0 M_{\eta}\left[r_0^2
  M_\pi^2\right] = 1.256(54)_\mathrm{stat}(100)_\mathrm{sys}$ and in 
physical units 
\begin{equation}
M_\eta(M_\pi) = 551(33)_\mathrm{stat}(44)_\mathrm{sys} \ \mathrm{MeV}  \ ,
\end{equation}
where the experimental mass-value of the neutral pion $M_{\pi^0} =
135\ \mathrm{MeV}$ has been used for $M_\pi$. In the $SU(2)$ chiral
limit we obtain $r_0M_\eta^0 =
1.230(65)_\mathrm{stat}(98)_\mathrm{sys}$ or $M_\eta^0 =
539(35)_\mathrm{stat}(43)_\mathrm{sys} \ \mathrm{MeV} \ .$ 

As the procedure used to correct the $\eta$ mass for mistuning of the
strange quark mass ignores a possible dependence on $\mu_\ell$,
$\mu_\sigma$, $\mu_\delta$ and $\beta$, it is desirable to have a
cross-check. In Ref.~\cite{Ottnad:2012fv} we discussed two possible options
for this. The first one is to study the ratio $M_\eta/M_K$, for which
it was shown that most of the strange quark dependence cancels. The
second possibility is to study the Gell-Mann-Okubo (GMO) relation which is motivated by
chiral perturbation theory. Here we simply repeat the analysis for both
cases, including the new data point for $D30.48$.  

For the ratio $(M_\eta/M_K)^2$  a linear extrapolation for all
available data in $(r_0M_\mathrm{PS})^2$ to the physical pion mass
point yields (see the shaded band in the right panel
of Figure~\ref{fig:r0meta}) $\left(M_\eta/M_\mathrm{K}\right)_{M_\pi}=
1.123(26) \ ,$ which agrees well with the experimental value
$\left(M_\eta/M_\mathrm{K}\right)_\mathrm{exp} = 1.100$. Using the
experimental value of $M_\mathrm{K^0}$ we obtain $M_\eta =
559(13)_\mathrm{stat}(45)_\mathrm{sys} \ \mathrm{MeV}\,.$ Note that in
this analysis the scale $r_0=0.45(2)\ \mathrm{fm}$ is only required
for determining the physical pion mass point. As the slope of the
extrapolation is rather small, the statistical uncertainty in $M_\eta$
is significantly smaller than for the direct extrapolation of
$(r_0M_\eta)^2$. 

Considering the GMO relation and again performing a linear
extrapolation in $(r_0M_\mathrm{PS})^2$ including all available data
we obtain $\left(3M_\eta^2/(4M_\mathrm{K}^2 - M_\pi^2)\right)_{M_\pi}
= 0.970(47)$ at the physical pion mass, which is in agreement with
experiment, $(3M_\eta^2/(4M_\mathrm{K}^2 - M_\pi^2))^\mathrm{exp} =
0.925$. Using the experimental values of $M_{\pi^0}$ and
$M_{\mathrm{K}^0}$ we now obtain $M_\eta =
561(14)_\mathrm{stat}(45)_\mathrm{sys}\ \mathrm{MeV}\,,$ where the
first error is statistical and the second systematic estimated again
from the scaling violations discussed above.

\begin{table}[t!]
 \centering
 \begin{tabular*}{0.9\textwidth}{@{\extracolsep{\fill}}lcccccc}
  \hline\hline
  ensemble & $aM_{\mathrm{PS}}$ & $aM_\mathrm{K}$ & $aM_\eta$ & $a M_{\eta'}$  \\
  \hline\hline
  $D30.48$   & $0.09776(45)$ & $0.17760(23)$ & $0.205(16)$ & $0.38(4)$ & \\
  $D45.32sc$ & $0.07981(30)$ & $0.17570(84)$ & $0.192(15)$ & $0.30(4)$ \\
  \hline\hline
  \vspace*{0.1cm}
 \end{tabular*}
 \caption{Results of $a M_\eta$, $a M_{\eta'}$ for ensembles $D30.48$
   and $D45.32sc$. and the corresponding values for the charged pion mass
   $M_{\mathrm{PS}}$ and the kaon mass $M_\mathrm{K}$. The $D30.48$ is
   new compared to Ref.~\cite{Ottnad:2012fv} and ensemble $D45.32sc$
   is needed for the mixed action results. Results for all the
   other ensembles can be found in Ref.~\cite{Ottnad:2012fv}.}
 \label{tab:masses}
\end{table}

\section{Results using OS strange and charm quarks}

In this section we will discuss first results for $\eta$ and $\eta'$
meson masses using OS strange and charm quarks. We focus here on one
ensemble, namely $D45.32sc$, see Table~\ref{tab:setup}.

For given values of $a\mu_\ell, a\mu_s$ and $a\mu_c$  we determine the
correlation matrix Eq.~(\ref{eq:corrmatrix})  
using the interpolating operators Eq.~(\ref{eq:Oplight}) and
Eq.~(\ref{eq:OSop}) using $900$ configurations. Two consecutive
configurations are spaced by $4$ trajectories of length
$1$. Disconnected contributions are estimated using $32$ Gaussian
volume sources per gauge configuration. We use local and fuzzed
operators and build 
correspondingly a $3\times 3$ or $6\times6$ correlation function matrix. Errors
are estimated using $1000$ bootstrap samples. The data is blocked in
blocks of length $5$ to account for autocorrelations.
Mass values of $\eta$ and $\eta'$ are extracted using the same methods
as in the unitary case.

\subsection{Matching The Strange Quark Mass}

\begin{figure}[t]
 \centering
 {\includegraphics[width=.48\linewidth]{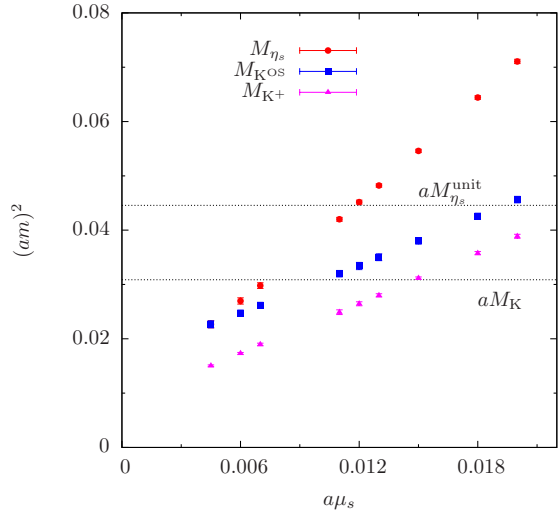}}
 \caption{We show $(M_{\eta_s})^2$, $(M_{K^\mathrm{OS}})^2$ and
   $(M_{K^+})^2$ extracted with OS-type valence quarks as functions of
   the bare OS strange quark mass $a\mu_s$. As horizontal lines we
   show the unitary values of $(aM_K)^2$ and
   $(aM^\mathrm{unit}_{\eta_s})^2$.}
 \label{fig:D45Match}
\end{figure}

As discussed in Section~\ref{sec:actions}, sea- and valence-actions can
be matched using various different observables. We will consider here 
$M_{\mathrm{K}^\mathrm{OS}}$, $M_{\mathrm{K}^+}$ and $M_{\eta_s}$. 
In Figure~\ref{fig:D45Match} we show
$(aM_{\mathrm{K}^\mathrm{OS}})^2$, $(aM_{\mathrm{K}^+})^2$ and
$(aM_{\eta_s})^2$ as functions of $a\mu_s$ for ensemble $D45.32sc$. We
also show both $(aM_{\eta_s}^\mathrm{unit})^2$ and
$(aM_\mathrm{K})^2$ as horizontal lines. The value of
$aM_\mathrm{K}$ can be found in
Table~\ref{tab:masses}. The value for $aM_{\eta_s}^\mathrm{unit}$ has
been determined using the connected contributions to the unitary
correlator matrix only, and its value is
$aM_{\eta_s}^\mathrm{unit}=0.2105(14)$.

From Figure~\ref{fig:D45Match} it is first of all clear that using
different valence quantities leads to very different matching values
for $a\mu_s$. Using for instance $M_{\mathrm{K}^+}$ leads to
$a\mu_s=0.0149(3)$, while matching $M_{\mathrm{K}^\mathrm{OS}}$
leads to $a\mu_s = 0.0102(3)$. Therefore, for this proceeding we
decided to use the following three values for the bare OS strange
quark mass
\[
a\mu_s = 0.01\,,\ 0.0125\,,\ 0.018\,,\ 0.025\,.
\]
These four values bracket the three matching values stemming from the
different matching quantities. It is worth noting that the leftmost
data points in Figure~\ref{fig:D45Match} correspond to the case where
$a\mu_s = a\mu_\ell$. Hence, $aM_{\mathrm{K}^\mathrm{OS}} =
aM_{\eta_s}$. The splitting in between $aM_{\mathrm{K}^\mathrm{OS}}$
and $aM_{\mathrm{K}^+}$ is an $\mathcal{O}(a^2)$ effect, which
disappears in the continuum limit.

As mentioned before we do not expect the precise charm quark mass
value to be important for $\eta$ and $\eta'$, which is also confirmed
by our unitary results. Therefore, we use only one value $a\mu_c =
0.172$ for the OS bare charm quark mass. It is obtained by matching
$aM_{\mathrm{K}^+} =aM_\mathrm{K}$ to determine
$a\mu_s$. $a\mu_c$ is then obtained via the relation Eq.~(\ref{eq:msc}).

\subsection{The valence strange quark mass dependence of the OS $\eta$ states}
\label{sec:sDep}

\begin{figure}[t]
  \centering
  \includegraphics[width=.48\linewidth]{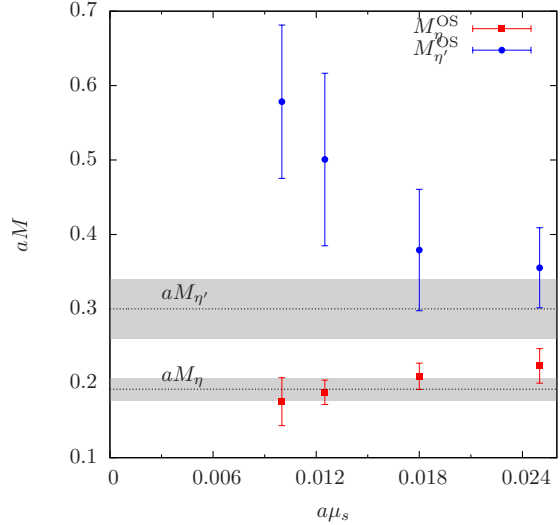} 
  \caption{$M_\eta^\mathrm{OS}$ and $M_{\eta'}^\mathrm{OS}$ as a
    function of $a\mu_s$ for $D45.32sc$. In addition we show as
    horizontal lines the corresponding unitary masses with errors as
    shaded band. Note that for $M_\eta$ and $M_{\eta'}$ we used higher statistics.}
  \label{fig:musdep}
\end{figure}

In Figure~\ref{fig:musdep} we show the lowest two states extracted
from a $6\times6$ matrix as a function of the OS strange quark mass
$a\mu_s$ for the $D45.32sc$ ensemble. The state with the lowest mass
should correspond to the 
$\eta$, the second to the $\eta'$ state. In addition to the OS
results we show as horizontal lines the unitary mass values discussed
in the previous section (see Table~\ref{tab:masses}). Note that the unitary values were produced with roughly twice the number of configurations but only $24$ instead of $32$ stochastic samples (cf. Table \ref{tab:setup}).

In general one observes from Figure~\ref{fig:musdep} that the $\eta$
mass value can be extracted with high statistical accuracy, while the
$\eta'$ suffers from similar noise as observed for the unitary
$\eta'$. In fact, for the $\eta$ meson mass we obtain slightly
better accuracy as compared to the unitary case. However, the noise
reduction trick applicable for OS strange and charm quarks does not
seem to help for the extraction of the OS $\eta'$ meson mass: in the
effective mass plots a plateau is only hardly visible. Hence, we
expect that the OS $\eta'$ is affected by similarly large systematic
uncertainties as the unitary $\eta'$.

Therefore, we consider in the following only the OS $\eta$ meson and
its $\mu_s$ dependence. It is interesting to understand which of the
matching observables discussed above yields the best agreement in
between OS and unitary $\eta$ mass values. First of all, we observe a
rather mild dependence of $M_\eta^\mathrm{OS}$ on the bare OS strange
quark mass $a\mu_s$, see Figure~\ref{fig:musdep}. In order to compare
to the unitary case, we compute 
\[
D_\eta^\mathrm{OS}\ \equiv\
\left[\frac{d(M_\eta^\mathrm{OS})^2}{d(M_{\mathrm{K}^+})^2}\right] =
 0.8(1)
\]
with statistical error only, and compare to $D_\eta=1.60(18)$ defined
in Eq.~\ref{eq:Deta}. The OS value is significantly smaller which we
attribute to the large sea quark contributions to the $\eta$ meson
mass.

As a consequence, the OS $\eta$ mass agrees within errors with the
unitary $\eta$ mass value for all $a\mu_s$ values considered. The
agreement is best around the $M_{\mathrm{K}^+}$ matching point,
though. Given the large uncertainties in the OS $\eta'$ meson mass, we
find also for the $\eta'$ meson masses at least marginal agreement
within errors. However, one should keep in mind the potentially large
systematics affecting the $\eta'$ mass determination.

\subsection{Comparing OS and Unitary Approach}

\begin{figure}[t]
  \centering
  \subfigure[]{\includegraphics[width=.48\linewidth]{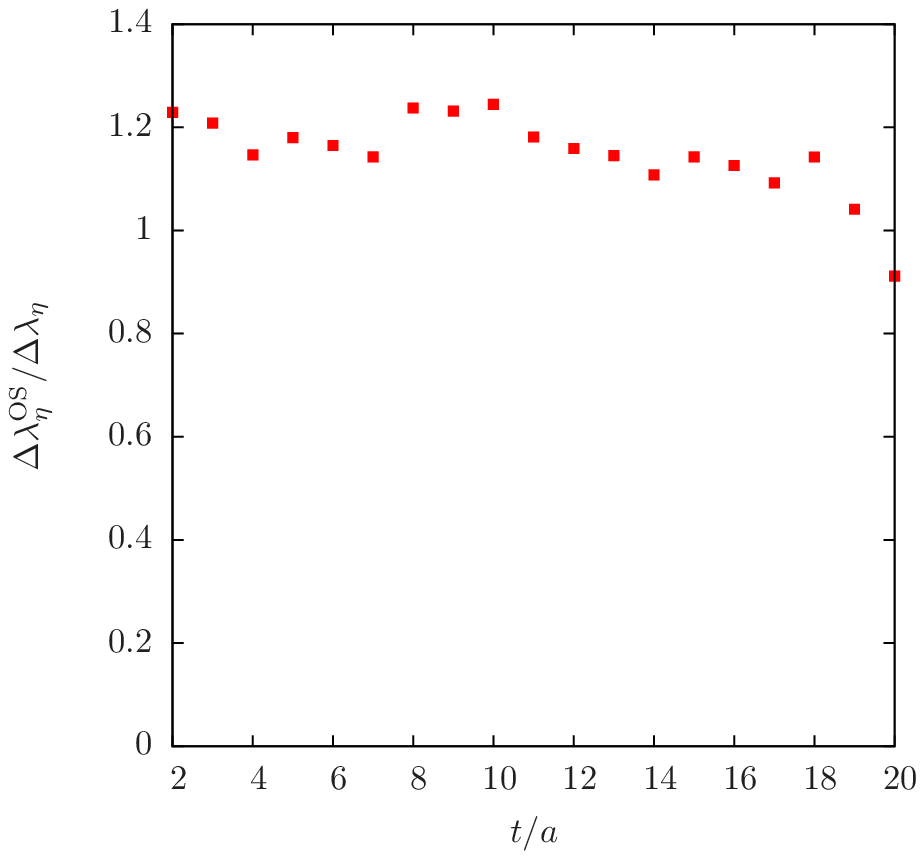}}\quad
  \subfigure[]{\includegraphics[width=.48\linewidth]{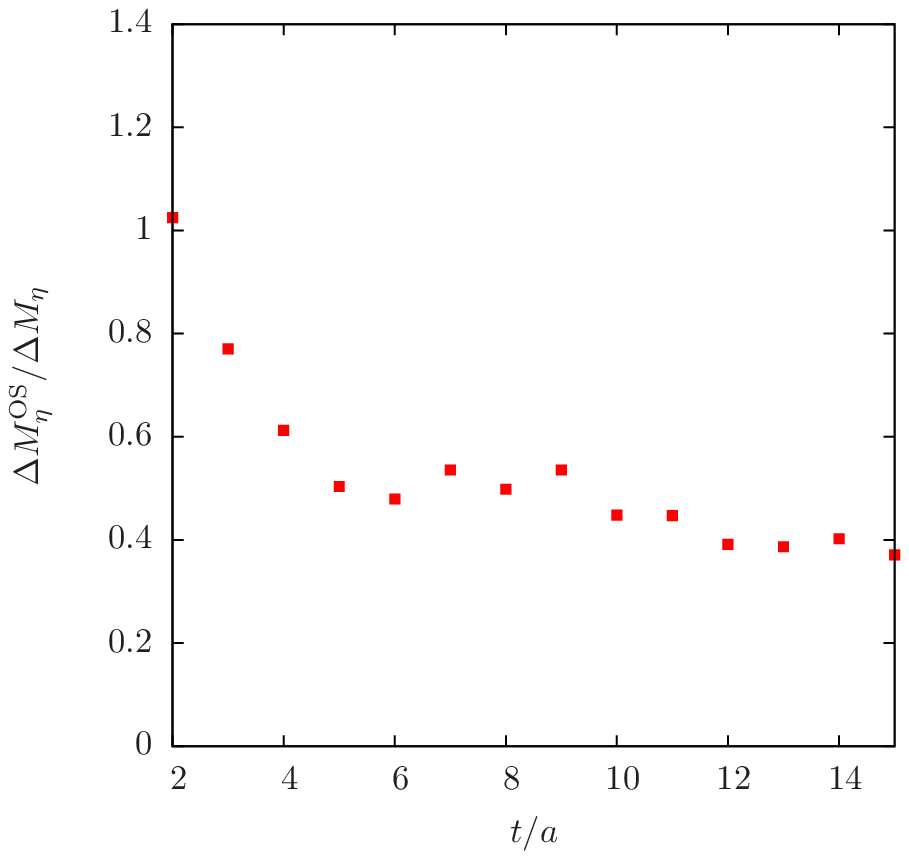}}
   \caption{(a) We show the OS to unitary ratio of relative errors of
     the $\eta$-eigenvalue as a function of $t/a$.
     (b) the same as (a) but for the effective masses.
   }
  \label{fig:compOS}
\end{figure}

It is also interesting to compare OS and unitary approach on a
correlator level. For this we computed the correlator matrix
Eq.~(\ref{eq:corrmatrix}) for $D45.32sc$ on the same set of
configurations with identical number of stochastic samples, namely
$N_s=24$. For both cases the GEVP is solved and the eigenvalues with
errors are extracted. We then compare in the left panel of
Figure~\ref{fig:compOS} the relative errors of the eigenvalues
corresponding to the $\eta$ state $\lambda_\eta$ by plotting the ratio
$\Delta\lambda^\mathrm{OS}_\eta/\Delta\lambda_\eta$ as a function of
$t/a$. This ratio is slightly larger or close to $1$. However, the
same ratio for the effective masses shown in the right panel of
Figure~\ref{fig:compOS} is from $t/a=5$ on equal or smaller than
$1/2$. 

This apparent contradiction is explained by a correlation between
$\lambda_\eta^\mathrm{OS}(t/a)$ and $\lambda_\eta^\mathrm{OS}(t/a+1)$
very close to one for all $t/a$ in the OS case. For the unitary case
the same correlation coefficient is smaller or equal to $0.8$. For
this reason the $\eta$ meson mass can be determined with slightly
better accuracy from the OS analysis than from the unitary one. The
reason is likely to be the variance reduction trick for OS strange
quarks, as the $\eta$ has a larger strange than light contribution.

\section{Summary and Outlook}

In this proceeding contribution we presented an update of our
investigation of $\eta$ and $\eta'$ meson properties using $N_f=2+1+1$
flavour of Wilson twisted mass fermions, first published
in Ref.~\cite{Ottnad:2012fv}. We added one additional ensemble at the
finest available lattice spacing. This allows us to estimate
systematics from the continuum extrapolation with higher
confidence. The final result for the $\eta$ meson mass at the physical
point stays virtually unchanged compared to Ref.~\cite{Ottnad:2012fv}
\[
M_\eta = 558(14)_\mathrm{stat}(45)_\mathrm{sys}\ \mathrm{MeV}\,.
\]
In addition we presented an exploratory investigation with a mixed
action, where valence strange and charm quarks are regularised as so
called Osterwalder-Seiler fermions. For one ensemble $D45.32sc$ we
matched valence and unitary actions using $M_\eta$. The obtained OS
bare strange quark mass is in good agreement to the one obtained using
kaons for matching. 

It turns out that in the OS approach the $\eta$ can be determined with
slightly higher statistical accuracy compared to the unitary case, when
the number of inversions is approximately matched. However, the
$\eta'$ is still noisy and we cannot determine it with higher accuracy
than in the unitary case. The OS $\eta$ mass shows a smaller
dependency on the OS kaon mass than the unitary $\eta$ mass on the
unitary kaon mass, which we attribute to significant sea quark
contributions to the $\eta$ meson. 

In a next step we plan to improve the $\eta'$ meson mass
determination. This could be reached by using the point-to-point
method described in Ref.~\cite{Jansen:2008wv} and/or by increasing 
our operator basis. Moreover, we will study the continuum
extrapolation of the OS $\eta$ meson mass for different matching
conditions.

\subsection*{Acknowledgements} 
We thank J.~Daldrop, E.~Gregory, B.~Kubis, C.~McNeile, 
U.-G.~Mei{\ss}ner, M.~Petschlies and M.~Wagner for useful
discussions. We thank U.~Wenger for his help in determining the Sommer
parameter. We thank the members of ETMC for the most enjoyable
collaboration. The computer time for this project was made available
to us by the John von Neumann-Institute for Computing (NIC) on the
JUDGE and Jugene systems in J{\"u}lich and the IDRIS (CNRS) computing
center in Orsay. In particular we thank U.-G.~Mei{\ss}ner for granting
us access on JUDGE. This project was funded by the DFG as a project in
the SFB/TR 16. Two of the authors (K. O. and C.U.) were supported by
the Bonn-Cologne Graduate School (BCGS) of Physics and Astronomie. K.C. was supported by Foundation for Polish Science fellowship "Kolumb". 
The open source software packages tmLQCD~\cite{Jansen:2009xp},
Lemon~\cite{Deuzeman:2011wz} and R~\cite{R:2005} have been used.

\bibliographystyle{h-physrev5}
\bibliography{bibliography}

\end{document}